**Engagement Phenotypes for a Sample of 102,684 AI Mental Health Chatbot Users and Dose-Response Associations with Clinical Outcomes**

Emma C. Wolfe,[1] Ting Su,[2] Olivier Tieleman,[2] Thomas D. Hull,[2] Matteo Malgaroli,[3] Caitlin A. Stamatis[2]

[1]University of Virginia, Department of Psychology, 485 McCormick Rd, Charlottesville, VA 22904

[2]Slingshot AI, 228 Park Ave S, PMB 679458, New York, NY 10003, US

[3]New York University School of Medicine, One Park Avenue, New York, NY 10016

**Corresponding Author:** Caitlin A. Stamatis, Ph.D.

**Email:** caitlin@slingshotai.com

**Funding Statement:** MM's research was supported by the National Institutes of Mental Health (NIMH) through grant K23MH134068, the National Center for Complementary and Integrative Health (NCCIH) through grant R34AT012943, and Slingshot AI. The funding sources had no role in the analysis, interpretation of the data, or decision to submit the manuscript for publication. The content is solely the responsibility of the authors and does not necessarily represent the official views of the NIH. ECW, TS, OT, TDH, and CAS have no external funding to declare.

**Competing Interests:** CAS, TS, OT, and TDH are employees of the company building the model and declare no non-financial competing interests. ECW and MM declare no competing interests.

**Abstract**

**Background**: Conversational AI chatbots are emerging as scalable mental health tools, but little is known about real world engagement or its relationship to clinical outcomes. **Objective**: To characterize engagement phenotypes among users of Ash, a purpose-built AI mental health chatbot, and examine associations with clinical change and working alliance. **Methods**: K-means clustering across eight behavioral features identified engagement phenotypes among 102,684 users. Subsamples completed the PHQ-9 (n=298), GAD-7 (n=298), and MSPSS (social support; n=194) baseline and 3 weeks; 11,437 users completed baseline Working Alliance Inventory (WAI). **Results**: Five engagement phenotypes emerged: Early Dropouts (52.2%), Power Users (1.6%), Intensive Users (4.1%), Weekly Users (25.3%), and a novel Concentrated User pattern (16.8%); across users, 66.9% had at least one overnight session (9pm-5am). Significant pre-post improvements occurred in depression ($d$ = -0.51), anxiety ($d$ = -0.57), and social support ($d$ = 0.22). An observed dose-response gradient in self-reported depression improvement was replicated in a larger sample using model-predicted PHQ-9 ($n$ = 23,813), with the largest improvements among high-engagement Power and Intensive Users ($d$ = -0.40 and -0.43) and the smallest among Early Dropouts ($d$ = -0.11). Higher working alliance predicted depression improvement and moderated the engagement-social support relationship. **Conclusions**: Engagement with AI mental health tools is multidimensional, and different clinical outcomes respond to different dimensions of use. Findings caution against treating session counts as a primary engagement metric and offer naturalistic evidence for the clinical value of purpose-built conversational AI.

## Introduction

Mental health disorders represent the leading cause of disability worldwide [1,2]. Despite the availability of evidence-based treatments, most affected individuals remain untreated due to systemic barriers including prohibitive financial costs, long waitlists, and a critical shortage of clinical professionals [3–7]. In the United States alone, 137 million people live in mental health professional shortage areas, and median wait times for psychiatric services can range from 43 to 67 days [8]. Digital mental health interventions (DMHIs) are one promising public health intervention, leveraging rapid increases to global smartphone access to disseminate discreet evidence-based support [9–11]. Among existing DMHIs, conversational artificial intelligence (AI) chatbots may represent uniquely scalable tools. Although experts have expressed safety and ethics concerns over use of general-purpose tools (e.g. ChatGPT, Grok AI) for mental health support, a growing body of evidence suggests that purpose-built tools (i.e. those designed specifically for emotional support) can disseminate safe, personalized, cost-effective mental health care. These tools, which incorporate evidence-based strategies and safety guardrails, have shown encouraging early benefit to user-reported anxiety, depression, and social support. [12–16]

Despite their potential, the effectiveness of purpose-built AI chatbots can be undermined by high attrition in research trials.[13,17] Users engage differently with these tools in the real world than in controlled laboratory studies, where participation is closely monitored and compensated. Efforts to understand engagement are further stymied by a lack of standardized metrics for engagement. Existing research frequently conflates usage time or simple interaction counts with meaningful use, which are largely measured from small-scale controlled trials.[18–20] As a result, the effect sizes in controlled trials may overestimate real-world benefit, and there is a need to examine usage patterns and outcomes in large, unselected samples.[18,20]

Recent work has shifted toward behavioral phenotyping as a more nuanced approach to characterizing engagement. Behavioral phenotyping uses unsupervised machine learning techniques (e.g. K-means clustering) to identify distinct engagement patterns, grouping users based on behavioral signatures including session frequency, duration, and depth of interaction rather than treating engagement as a single variable.[21,22] This person-centered methodology seeks to identify *what works for whom* by recognizing that different users may benefit from different patterns of use. For example, emerging evidence indicates that "efficient engagers" (i.e., users with fewer sessions but high affective and cognitive engagement) achieve outcomes comparable to or exceeding those of higher-volume users.[23] Other studies have identified a range of engagement profiles across DMHIs, including light, deep, and time-limited users, suggesting that the relationship between engagement quantity and clinical benefit is not straightforward.[19,22–24] Whether high-volume engagement yields diminishing returns, as some preliminary evidence in the context of AI chatbots suggests,[13,25,26] remains an open question that large-scale, real-world data are well positioned to address.

Behavioral engagement patterns, however, capture only one dimension of the user-chatbot interaction. How users *experience* their relationship with a conversational AI (i.e., the degree to which they perceive trust, collaboration, and shared purpose) may shape both how they engage and whether they benefit. This relational dimension is captured in the concept of the digital working alliance (WAI). In face-to-face therapy, the therapeutic bond between client and provider is one of the strongest and most consistent predictors of treatment outcome.[27] Preliminary evidence suggests that users of conversational AI can form relationships characterized by empathy, trust, and perceived collaboration that are at minimum, non-inferior to those formed with human support.[28–30] There is an existing, but limited, body of evidence suggesting an association between engagement patterns and degree of therapeutic alliance with human and AI-enabled digital interventions.[31] Understanding the nature of this alliance is particularly important given that a substantial proportion of chatbot use occurs during overnight hours, when human support is largely unavailable. As digital working alliance is a relatively new construct in the context of mental health AI tools and given concerns about the overly affirming nature of conversational AI for emotional support,[32,33] further characterization of its relationship to both engagement patterns and clinical outcomes is needed.

The current study addresses these gaps by providing a large-scale, longitudinal analysis of real-world engagement and clinical outcomes with Ash, a conversational AI designed specifically for mental health support. Rather than characterizing engagement as "high" or "low", we operationalize it as a multidimensional behavioral construct measured across eight features, capturing session frequency, duration, depth, consistency, and temporal patterns of use. Utilizing a large naturalistic sample ($n$=102,684 for behavioral clustering; subsamples of $n$=194-298 users with clinical outcome data), we present the following aims and hypotheses:

*Aim 1* is to characterize behavioral phenotypes of engagement using unsupervised machine learning to identify distinct behavioral phenotypes. Consistent with emerging research [19,23,24], we hypothesize that most users will exhibit episodic engagement, with a smaller subset characterized by sustained, high-volume use (**H1**). *Aim 2* is to test whether engagement with Ash is associated with clinical improvement in depression, anxiety, and perceived social support. Given effect sizes observed in prior research on Ash[15], we hypothesize significant improvements across all three outcomes (**H2**) *Aim 3* is to examine the dose-response relationship between engagement and clinical outcomes. We hypothesize that higher-engagement clusters will evidence larger clinical gains (**H3**). Finally, as an *Exploratory Aim*, we examine the association between perceived therapeutic alliance with the AI and both engagement patterns and clinical outcomes.

## Methods

### Participants and Procedures

Participants were users of Ash (talktoash.com), a purpose-built conversational AI tool for mental wellbeing. After viewing the app Terms and Conditions in the process of onboarding, users provided opt-in consent for the use of their de-identified data for research purposes. The study was approved by the Institutional Review Board at the NYU School of Medicine (i25-01177). Research was conducted in accordance with the Declaration of Helsinki. The full sample for engagement clustering analysis (*n* = 102,684) consisted of a random selection of real-world users from January 2025 - January 2026 who had at least one session recorded in the production database. No exclusion criteria were applied to the behavioral analysis sample beyond the requirement of at least one recorded session and opt-in consent for the use of their data.

In-app clinical assessments were administered to randomly selected subsamples of users at baseline and week 3. Three subsamples with complete data at both timepoints were included in the present analyses, with zero users shared across groups. Sample 1 (*n1* = 298) was administered the PHQ-9, Sample 2 (*n2* = 298) was administered the GAD-7, and Sample 3 (*n3* = 194) was administered the MSPSS. In all three samples, the individual measure was administered at two timepoints approximately three weeks apart. See Table 1 for demographic details about each sample.

**Table 1.** *Characteristics of Clinical Subsamples*

| | Sample 1 (PHQ-9; *n* = 298) *n* (%) | Sample 2 (GAD-7; *n* = 298) *n* (%) | Sample 3 (MSPSS; *n* = 194) *n* (%) |
|---|---|---|---|
| *Age* | | | |
| 18-24 | 65 (21.8%) | 52 (17.4%) | 28 (14.4%) |
| 25-34 | 71 (23.8%) | 81 (27.2%) | 63 (32.5%) |
| 35-44 | 80 (26.8%) | 87 (29.2%) | 59 (30.4%) |
| 45-54 | 49 (16.4%) | 41 (13.8%) | 33 (17.0%) |
| 55+ | 21 (7.0%) | 33 (11.1%) | 8 (4.1%) |
| Prefer not to say | 12 (4.0%) | 4 (1.3%) | 3 (1.5%) |
| *Gender* | | | |
| Man | 84 (28.2%) | 81 (27.2%) | 59 (30.4%) |
| Woman | 201 (67.4%) | 207 (69.5%) | 115 (59.3%) |
| Nonbinary/Other | 10 (3.4%) | 7 (2.3%) | 6 (3.1%) |
| Missing | 3 (1.0%) | 3 (1.0%) | 14 (7.2%) |
| *Race/ethnicity* | | | |
| White | 191 (64.1%) | 189 (63.4%) | 123 (63.4%) |
| Asian | 22 (7.4%) | 33 (11.1%) | 26 (13.3%) |
| Hispanic/Latino | 8 (2.7%) | 10 (3.4%) | 9 (4.6%) |
| Black/African American | 32 (10.7%) | 17 (5.7%) | 10 (5.2%) |
| Multiracial | 21 (7.0%) | 27 (9.1%) | 6 (3.1%) |
| Other | 7 (2.3%) | 11 (3.7%) | 7 (3.6%) |

**Note.** PHQ-9 = Patient Health Questionnaire-9; GAD-7 = Generalized Anxiety Disorder-7; MSPSS = Multidimensional Scale of Perceived Social Support. Percentages are calculated within each sample.

The behavioral and clinical samples overlap but are not identical: 74 users with clinical data were not present in the engagement features dataset (e.g., due to data pipeline differences or session-level filtering), and 4,860 users had Working Alliance Inventory (WAI) data but were not included in the main clustering sample.

## Materials & Measures

### *Clinical Outcomes*

**Patient Health Questionnaire - 9 (PHQ-9).** Depression symptoms were assessed using the Patient Health Questionnaire-9 (PHQ-9;[34]), a 9-item scale with scores ranging 0-27. Higher scores indicate greater symptom severity, with established cut points at 5 (mild), 10 (moderate), 15 (moderately severe), and 20 (severe). The PHQ-9 has demonstrated strong internal consistency in other studies [35,36] and within this sample (Cronbach's *a* = .845-.891). The PHQ-9 was administered to a randomly selected subsample of users (*n* = 298).

**Generalized Anxiety Disorder -7 (GAD-7).** Anxiety symptoms were assessed with the Generalized Anxiety Disorder-7 (GAD-7;[37]), a 7-item scale with scores ranging 0-21. Higher scores indicate greater anxiety. The GAD-7 has demonstrated strong internal consistency in other studies [38,39] and within this sample (Cronbach's *a* = .860-.880). The GAD-7 was administered to a separate, randomly selected subsample of users (*n* = 298).

**Multidimensional Scale of Perceived Social Suppor**t (MSPSS). Social support was assessed with the Multidimensional Scale of Perceived Social Support (MSPSS; [40]), an 8-item scale tapping perceived support from family, friends, and significant others (scores 8-56; higher scores equate to more support). The MSPSS has demonstrated strong internal consistency in other studies [41,42] and within this sample (Cronbach's *a* = .782-.853). The MSPSS was administered to a third, randomly selected subsample of users (*n* = 194).

***Working Alliance***

**Working Alliance Inventory (WAI).** Therapeutic alliance was measured using the Working Alliance Inventory (WAI; [43], modified for human-AI interaction. Scores range 12-60 (higher = stronger perceived alliance). WAI was completed at a single timepoint near the start of engagement. The WAI has demonstrated strong internal consistency in other studies [43,44] and within this sample (Cronbach's *a* = .954). WAI data were available for a subsample of 11,437 users (11.1% of the full behavioral sample) spanning the three clinical subsamples and additional users who did not complete any other outcome measure. Clinical analyses involving WAI therefore use overlapping subsamples ($n$ = 209 for PHQ-9, $n$ = 207 for GAD-7, and $n$ =133 for MSPSS).

## Analytic Approach

### *Behavioral Engagement Feature Extraction.*

Eight behavioral engagement features were extracted from production session logs at the user level: (1) total session count, (2) active span in days (first to last session), (3) sessions per week (session count divided by active span in weeks), (4) average session duration in minutes, (5) average user messages per session, (6) usage trend (slope of session frequency over time), (7) day density (proportion of days in active span with at least one session), and (8) proportion of sessions occurring between 9 PM and 5 AM (night usage). These features were selected to capture a multidimensional representation of user engagement behavior, consistent with best practices described in previous work [20,23,45], and entered into K-means clustering analysis (see details below).

### *Clinical Outcomes Analyses.*

Pre-to-post change was computed as T2 minus T1 for each outcome (PHQ-9, GAD-7, MSPSS). Overall changes were tested using paired-samples t-tests. Effect sizes were computed as Cohen's d using the pre-post standard deviation.

### *Engagement and Outcomes.*

Between-cluster differences in mean change scores (PHQ-9, GAD-7, MSPSS) were tested using one-way ANOVA across the five clusters. Per-cluster within-group changes were also tested using paired t-tests. To obtain an estimate of clinical outcome change as related to engagement clusters in a larger and more representative sample, we replicated these analyses using PHQ-9 scores predicted by a five-model machine-learning ensemble in 23,813 users with two or more prediction windows (see Supplementary Materials for details on PHQ-9 score prediction). We examined change across two time horizons (first-to-last prediction window and a fixed four-week interval) and, as a sensitivity analysis, repeated each within a high-confidence subset (users above the median model prediction confidence). As a sensitivity analysis, we also examined associations between individual engagement features and clinical outcomes; for details, see Supplementary Materials.

***Working Alliance, Engagement, and Outcomes.***

WAI differences across clusters were tested with one-way ANOVA (F-test) followed by inspection of pairwise differences. Given the exploratory nature of this hypothesis, simple associations between WAI and engagement features were assessed using Pearson correlations in the full WAI sample. The relationship between baseline WAI and clinical change was tested using ordinary least squares (OLS) regression in the clinical subsample with WAI data, controlling for baseline symptom severity. The potential moderating role of WAI (WAI × engagement interaction) was tested by adding a product term to the OLS regression model.

**Data Availability**

The datasets generated and/or analyzed for this study are not publicly available for proprietary reasons but may be made available to qualified researchers on reasonable request from the corresponding author.

**Code Availability**

The underlying code for this study is not publicly available but may be made available to qualified researchers on reasonable request from the corresponding author.

## Results

### Aim 1: Identification of Engagement Clusters

A total of 102,684 users were included in the behavioral clustering analysis. Prior to behavioral clustering, users with 1 or 2 total sessions were identified as an Early Dropout group (C0; *n* = 53,573; 52.2%). K-means clustering was applied to the remaining 49,111 users with 3 or more sessions, evaluated across k = 2-10 cluster solutions. The eight engagement features were z-score standardized prior to clustering. Cluster selection used two criteria: silhouette score (measuring within-cluster cohesion relative to between-cluster separation) with higher scores indicating better separation and cluster stability (proportion of users assigned to the same cluster across two independent runs with different random seeds). K = 4 was selected as optimal, yielding the highest silhouette score across all tested solutions (0.233) and 99.9% stability. The final five-cluster solution (C0-C4) thus combines the Early Dropout group (C0) with the four k-means clusters (C1-C4). The full five-cluster solution (including C0) is shown in Table 2 and a figure depiction is available in the Supplementary Materials (Supp. Figure 1).

The four k-means clusters showed distinct and interpretable engagement profiles beyond the two we hypothesized would emerge ("power users" and "weekly users"). As hypothesized, one cluster (Cluster 1: Power Users; *n* = 1,611; 1.6%) was characterized by particularly high session count (*M* = 102.1) and active span (*M* = 132.2 days), representing sustained long-term use. In addition to this expected cluster, three unique user clusters emerged. Cluster 2 (Intensive Users; *n* = 4,201; 4.1%) was distinguished by the longest average session duration (*M* = 92.0 min) and highest message density (*M* = 74.8 messages/session), reflecting intensive depth of engagement within each session over a 24.7-day active span. Cluster 3 (Weekly Users; *n* = 26,016; 25.3%) showed infrequent use spread over extended periods (*M* = 72.4 days, 1.3 sessions/week), consistent with episodic or as-needed engagement. Cluster 4 (Concentrated Users; *n* = 17,283; 16.8%) showed moderately frequent use concentrated in a short window (*M* = 8.9 days, 4.5 sessions/week), consistent with an intensive but time-limited burst of engagement.

**Table 2.** *Engagement Cluster Profiles (n = 102,684)*

| Cluster | n (%) | Sessions | Active span | Sessions /week | Session duration (min) | Messages /session | Night sessions (%) |
|---|---|---|---|---|---|---|---|
| | | *M (SD)* | *M (SD)* | *M (SD)* | *M (SD)* | *M (SD)* | *M (SD)* |
| C0: Early dropouts | 53,573 (52.2%) | 1.3 (0.5) | 4.4 (15.1) | 1.2 (0.5) | 28.1 (36.8) | 24.7 (31.8) | 1.3 (0.5) |
| C1: Power users | 1,611 (1.6%) | 102.1 (55.4) | 132.2 (61.1) | 6.4 (3.6) | 50.1 (25.8) | 31.4 (20.2) | 102.1 (55.4) |
| C2: Intensive users | 4,201 (4.1%) | 11.3 (11.9) | 24.7 (32.0) | 4.4 (3.3) | 92.0 (44.0) | 74.8 (44.0) | 11.3 (11.9) |
| C3: Weekly users | 26,016 (25.3%) | 10.7 (9.9) | 72.4 (53.7) | 1.3 (0.9) | 27.2 (17.6) | 21.2 (13.8) | 10.7 (9.9) |
| C4: Concentrated users | 17,283 (16.8%) | 7.4 (7.6) | 8.9 (10.2) | 4.5 (2.1) | 28.3 (19.0) | 19.6 (12.5) | 7.4 (7.6) |

**Note.** *n* = 102,684. Sessions/Week calculated over active span. Night = 9 PM-5 AM local time. Day Density = % of active span days with at least one session.

Temporal patterns of use were the only engagement feature not associated with distinct clusters. Across all users and all clusters, 66.9% of users had at least one session occurring between 9 PM and 5 AM, and on average 38.6% of a given user's sessions took place during these overnight hours.

**Aim 2: Overall Clinical Outcomes**

Depression symptoms (PHQ-9) decreased by 3.08 points on average in the depression subsample (*n* = 298; *SD* = 6.04; t(297) = -8.81; *d* = -0.51; *p* < .001). Anxiety symptoms (GAD-7) decreased by 2.79 points in the anxiety subsample (*n* = 298; *SD* = 4.91; t(297) = -9.80; *d* = -0.57; *p* < .001). In the MSPSS subsample (*n* = 194), a significant improvement in perceived social support was observed (MSPSS: +1.57 points; t(193) = 3.05; *d* = +0.22; *p* = .002). Baseline-to-Week-3 change can be found in the Supplementary Materials (Supp. Table 1).

**Figure 1.** *Overall Pre-Post Clinical Outcomes*

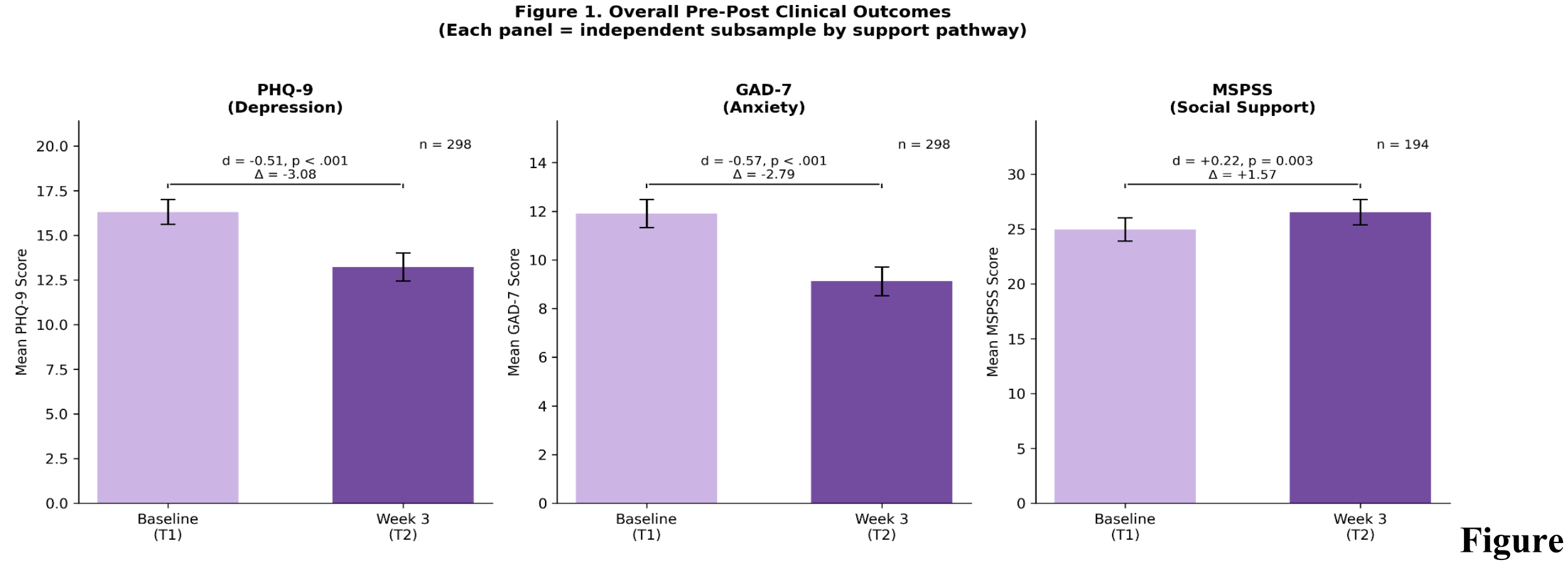

**Figure**

**Figure 1 Caption:** Mean pre-post scores. Left: PHQ-9 (*n* = 298). Center: GAD-7 (*n* = 298). Right: MSPSS (*n* = 194). Each panel represents an independent, non-overlapping subsample. Error bars = 95% CI. Cohen's *d* and *p*-values shown above brackets.

## Aim 3: Engagement Dose-Response

### *Clinical Outcomes by Cluster*

On the PHQ-9 (Table 3; Figure 2), all clusters showed improvement, with effect sizes ranging from *d* = -0.31 (C0: Early Dropouts) to *d* = -0.67 (C1: Power Users). Within-cluster improvements were statistically significant for Power Users (C1; *p* < .001) and Weekly Users (C3; *p* < .001). For GAD-7, Power Users (C1; *d* = -0.64, *p* < .001), Weekly Users (C3; *d* = -0.56, *p* < .001), and Concentrated Users (C4; *d* = -0.71, *p* = .002) all showed significant within-cluster improvement. For social support (MSPSS), the Power Users cluster (C1) showed the largest and only statistically significant improvement (change = +5.12 points; *d* = +0.50; *p* = .022), while other clusters showed smaller and non-significant changes. All between-cluster ANOVAs were non-significant (*p* = .089).

**Figure 2. Clinical Outcome Effect Sizes by Cluster**

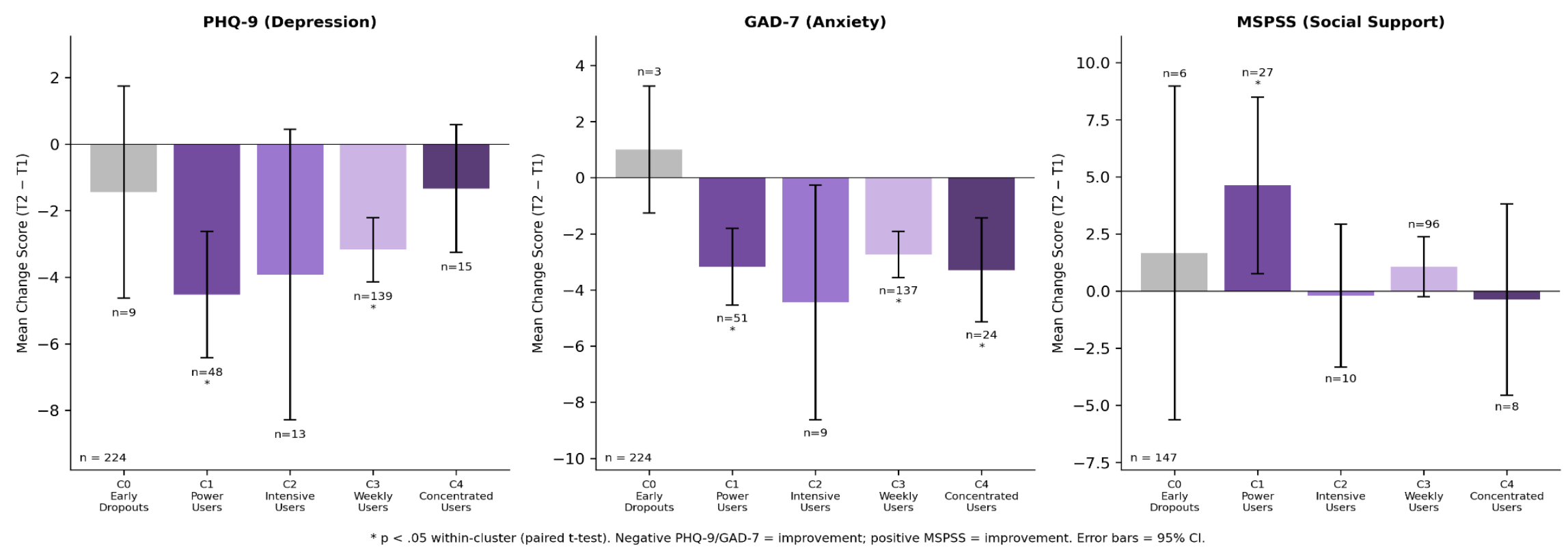

**Figure 2 Caption:** Left: PHQ-9 Depression ($n$ = 224). Center: GAD-7 Anxiety ($n$ = 224). Right: MSPSS Social Support ($n$ = 147). Negative PHQ-9/GAD-7 values indicate symptom improvement; positive MSPSS values indicate improvement. Cluster $n$ shown above each bar. *$p$ < .05 within-cluster (paired t-test).

**Table 3.** *PHQ-9, GAD-7, and MSPSS Outcomes by Engagement Cluster*

| Cluster | *n* | Baseline *M* (*SD*) | Week 3 *M* (*SD*) | Change *M* (*SD*) | Cohen's *d* | *p* |
|---|---|---|---|---|---|---|
| *PHQ-9 (Depression)* — ANOVA across clusters: $F(4, 219) = 0.88$, $p = .488$ | | | | | | |
| C0: Early dropouts | 9 | 17.6 (5.0) | 16.1 (7.0) | -1.44 (4.6) | -0.31 | .400 |
| C1: Power users | 48 | 16.8 (5.8) | 12.3 (7.0) | -4.52 (6.7) | -0.67 | < .001 |
| C2: Intensive users | 13 | 16.3 (6.6) | 12.4 (7.0) | -3.92 (7.7) | -0.49 | .056 |
| C3: Weekly users | 139 | 16.4 (6.1) | 13.2 (6.8) | -3.17 (5.7) | -0.55 | < .001 |
| C4: Concentrated users | 15 | 15.2 (6.6) | 13.9 (7.3) | -1.33 (4.2) | -0.35 | .068 |
| *GAD-7 (Anxiety)* — ANOVA across clusters: $F(4, 219) = 0.67$, $p = .608$ | | | | | | |
| C0: Early dropouts | 3 | 13.3 (2.5) | 14.3 (2.1) | +1.00 (1.7) | +0.61 | .478 |
| C1: Power users | 51 | 12.3 (4.8) | 9.2 (5.3) | -3.18 (5.0) | -0.64 | < .001 |
| C2: Intensive users | 9 | 12.0 (4.9) | 7.6 (5.0) | -4.44 (6.4) | -0.69 | .072 |
| C3: Weekly users | 137 | 12.0 (5.1) | 9.3 (5.2) | -2.74 (4.8) | -0.56 | < .001 |
| C4: Concentrated users | 24 | 11.8 (5.4) | 8.5 (5.0) | -3.29 (4.5) | -0.71 | .002 |
| *MSPSS (Social Support)* — ANOVA across clusters: $p = .089$ | | | | | | |
| C0: Early dropouts | 6 | 26.8 (7.3) | 28.5 (9.4) | +1.67 (9.1) | +0.20 | .674 |
| C1: Power users | 25 | 24.8 (7.6) | 26.4 (8.2) | +5.12 (10.5) | +0.50 | .022 |
| C2: Intensive users | 15 | 25.7 (6.8) | 24.9 (7.5) | -0.80 (4.35) | -0.19 | .488 |
| C3: Weekly users | 93 | 24.6 (7.4) | 25.8 (7.9) | +1.17 (6.66) | +0.18 | .093 |
| C4: Concentrated users | 8 | 26.0 (5.9) | 25.6 (8.4) | -0.38 (6.05) | -0.07 | .866 |

**Note.** PHQ-9 = Patient Health Questionnaire-9; GAD-7 = Generalized Anxiety Disorder-7; MSPSS = Multidimensional Scale of Perceived Social Support. *M* = mean; *SD* = standard deviation. Cohen's *d* is the within-cluster standardized change score (negative values indicate symptom reduction for PHQ-9 and GAD-7 and reduced perceived support for MSPSS). Cluster-level *p* values are from paired-samples *t* tests of baseline-to-Week 3 change.

**Figure 3.** Cohen's *d* by engagement cluster for PHQ-9, GAD-7, and MSPSS

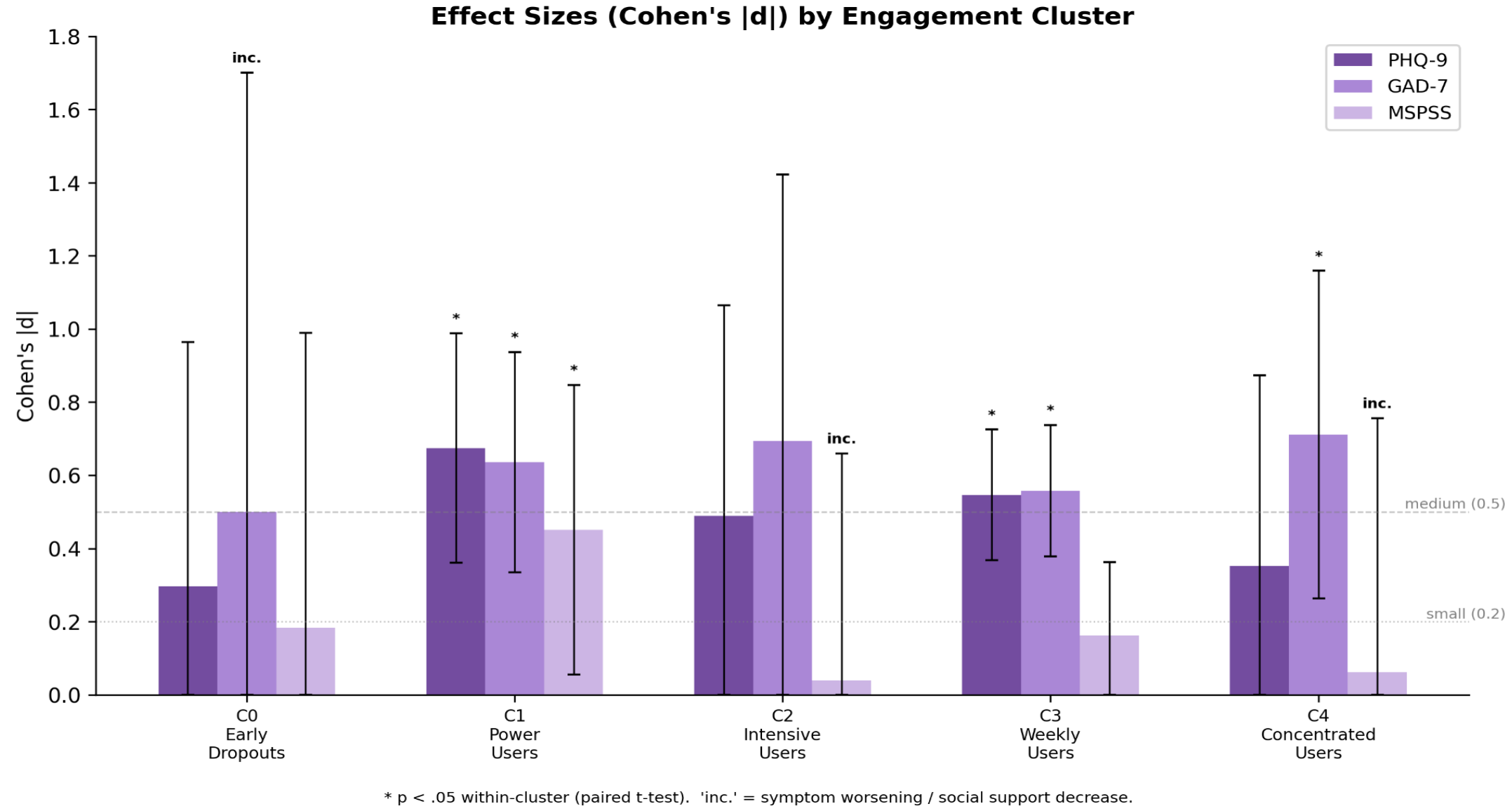

**Figure 3 Caption:** $p < .05$ within-cluster (paired t-test). 'inc.' indicates symptom worsening (C0 GAD-7) or social support decrease (C2, C4 MSPSS). Dashed/dotted lines mark small (0.2) and medium (0.5) effect size benchmarks.

**Replication in Predicted PHQ-9 Sample (*n*=23,813).** All clusters showed statistically significant reductions in predicted PHQ-9 from first to last prediction window (all $p < .001$; Figure 4; Supp. Table 2). A one-way ANOVA indicated significant between-cluster differences in predicted PHQ-9 change ($F(4, 23{,}808) = 31.49$, $p < .0001$). The two highest-engagement phenotypes showed the largest improvements—Intensive Users ($M = -2.08$, $SD = 4.84$, $d = -0.43$) and Power Users ($M =$

-2.06, *SD* = 5.13, *d* = -0.40), which did not differ from one another ($p$ = 1.00)—followed by Concentrated Users (*M* = -1.69, *SD* = 4.99, *d* = -0.34), Weekly Users (*M* = -1.16, *SD* = 5.06, *d* = -0.23), and Early Dropouts (*M* = -0.55, *SD* = 4.99, *d* = -0.11). The gradient was robust across analyses, holding in a high-prediction-confidence subset ($n$ = 12,122; $F(4, 12{,}117) = 12.84$, $p < .0001$) and over a fixed four-week interval ($n$ = 9,282; $F(4, 9{,}277) = 7.05$, $p < .0001$; Supp. Table 3).

**Figure 4.** Change in predicted PHQ-9 score by cluster (*n*=23,813).

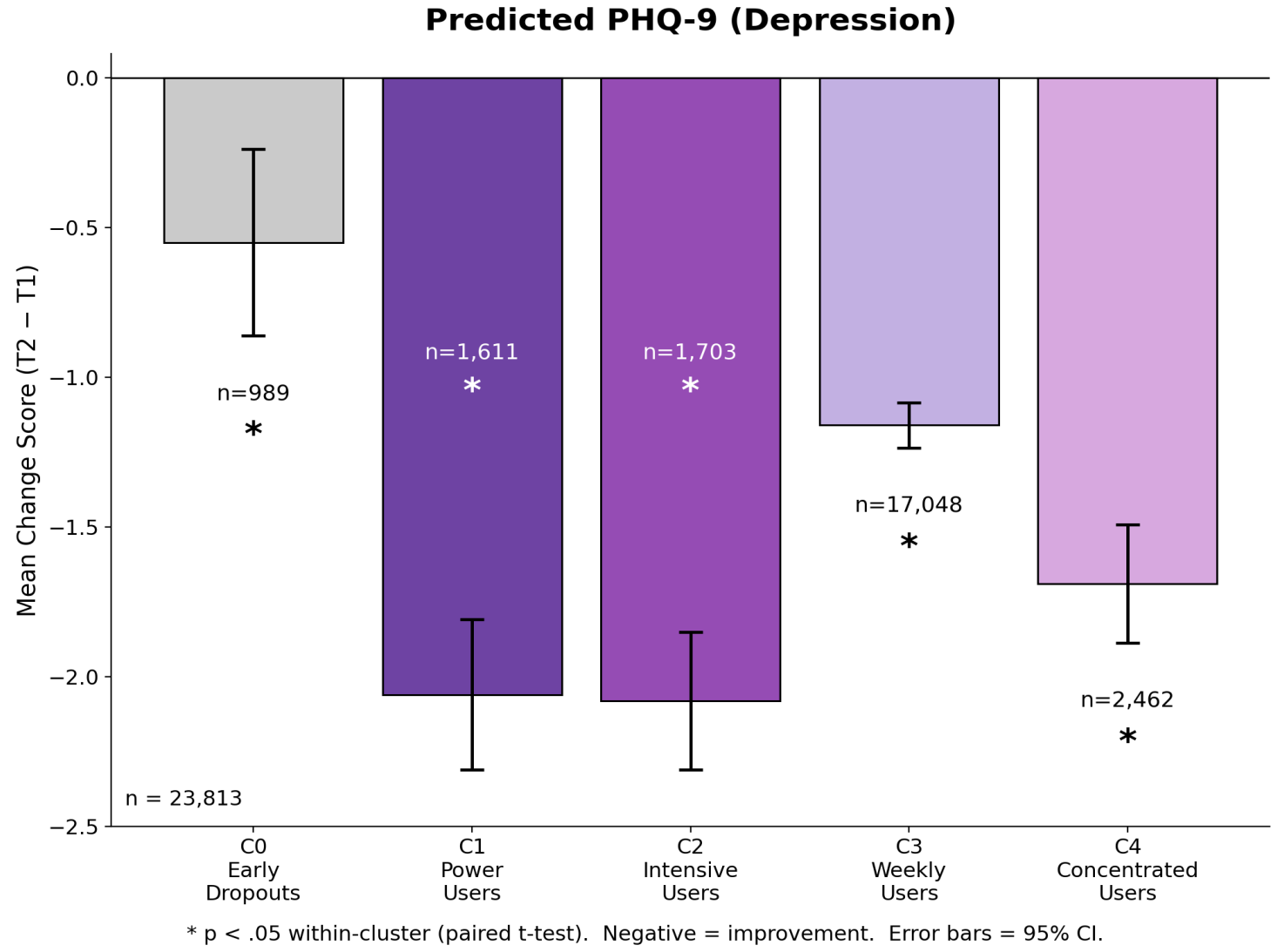

### ***Sensitivity Analysis: 21-Day Engagement Features as Predictors of Self-Reported Clinical Change***

Controlling for baseline PHQ-9, significantly lower week-3 PHQ-9 scores were associated with total message volume ($\beta = -0.14$, $SE = 0.05$, $sr^2 = 0.031$, $p = 0.009$), average messages per session ($\beta = -0.12$, $SE = 0.05$, $sr^2 = 0.023$, $p = 0.022$), and total active messaging time ($\beta = -0.12$, $SE = 0.05$, $sr^2 = 0.021$, $p = 0.029$). No 21-day engagement features significantly predicted week-3 GAD-7 scores. Controlling for baseline MSPSS, significantly higher week-3 MSPSS scores were associated with total active messaging time ($\beta = 0.17$, $SE = 0.06$, $sr^2 = 0.041$, $p = 0.005$) and session count ($\beta = 0.12$, $SE = 0.06$, $sr^2 = 0.021$, $p = 0.045$). Total messages ($\beta = 0.11$, $p = 0.061$) and active span days ($\beta = 0.10$, $p = 0.082$) showed trending effects. See Supplementary Results, Supp. Table 4, and Supp. Figure 2 for full results.

## Exploratory Aim: Digital Working Alliance

### *WAI Across Engagement Clusters*

Working Alliance Inventory (WAI) data were available for 11,437 users (11.1% of the full sample). Baseline WAI scores differed significantly across engagement clusters ($F = 16.59$, $p < .0001$; Table 4; Figure 5). Power Users (C1) reported the highest WAI ($M = 45.1$, $SD = 11.2$), followed by Intensive Users (C2; $M = 44.4$) and Weekly Users (C3; $M = 44.3$). Early Dropouts (C0; $M = 42.2$) and Concentrated Users (C4; $M = 42.4$) reported the lowest WAI scores. Correlational analyses in the full WAI sample indicated small but significant positive associations between WAI and engagement: session count ($r = 0.056$, $p < .001$), active span ($r = 0.077$, $p < .001$), and session duration ($r = 0.036$, $p < .001$).

**Table 4.** *Working Alliance Inventory (WAI) by Engagement Cluster (n = 11,437)*

| Cluster | *n* | Mean (*SD*) | Median | Range |
|---|---|---|---|---|
| C0: Early dropouts | 496 | 42.2(12.5) | 43 | 12-60 |
| C1: Power users | 1249 | 45.1 (11.2) | 47 | 12-60 |
| C2: Intensive users | 727 | 44.4 (12.5) | 47 | 12-60 |
| C3: Weekly users | 7009 | 44.3 (11.5) | 46 | 12-60 |
| C4: Concentrated users | 1956 | 42.4 (12.1) | 44 | 12-60 |

### *WAI as Predictor of Clinical Outcomes*

**Working Alliance and Clinical Outcomes.** Higher baseline WAI significantly predicted lower week-3 PHQ-9 scores after controlling for baseline severity ($n = 209$, $\beta = -0.13$, $SE = 0.06$, $sr^2 = 0.027$, $p = 0.018$). WAI did not significantly predict week-3 GAD-7 ($n = 207$, $\beta = -0.09$, $SE = 0.06$, $p = 0.122$). For social support, higher WAI significantly predicted higher week-3 MSPSS scores ($n = 133$, $\beta = 0.18$, $SE = 0.07$, $sr^2 = 0.045$, $p = 0.014$; Supp. Table 5).

**Figure 5.** *Mean WAI scores by engagement cluster.*

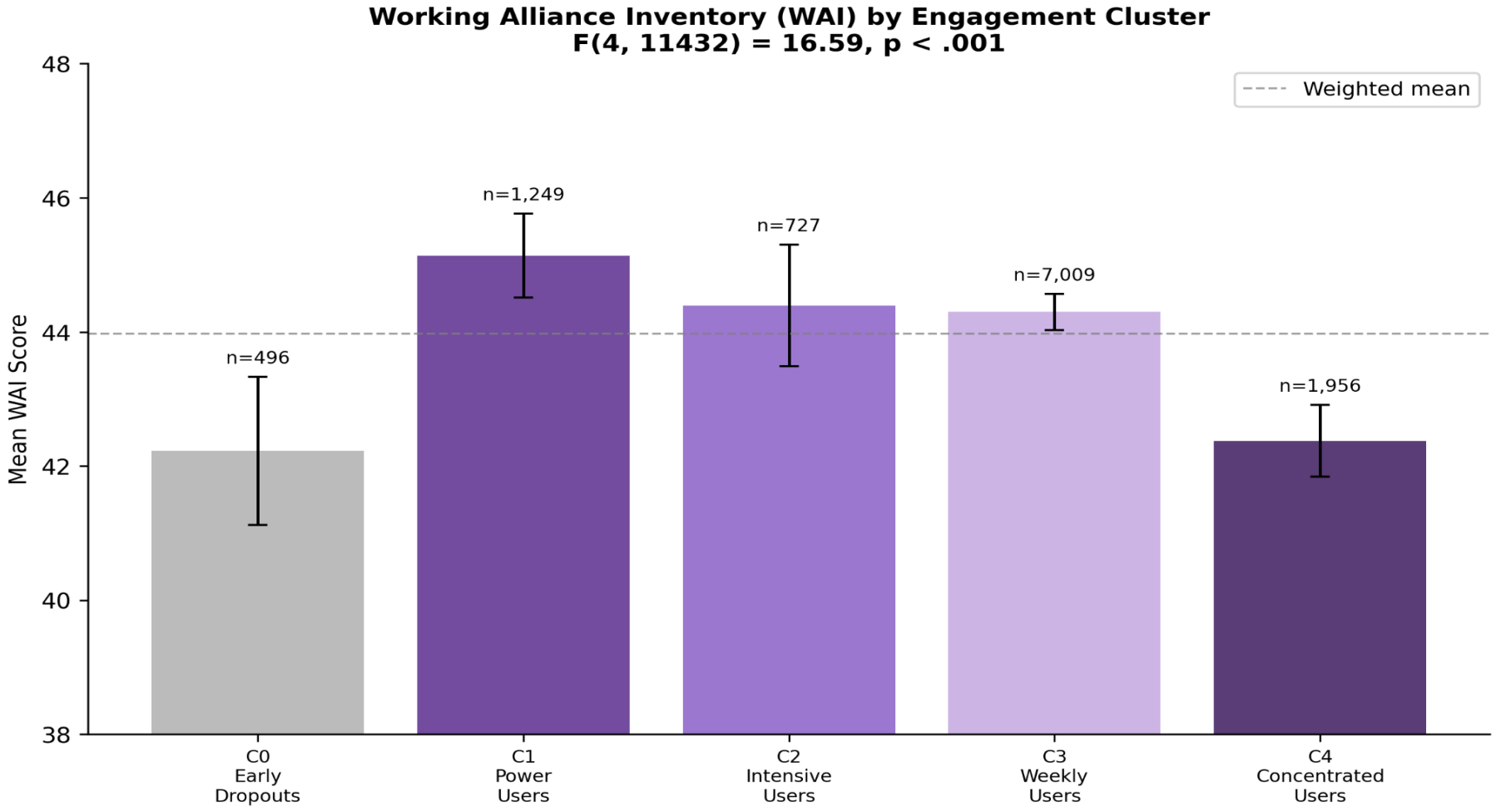

**Figure 5 Caption:** Error bars = 95% CI. Dashed line = weighted sample mean. $F(4, 11432) = 17.48$, $p < .001$.

**Moderation by Working Alliance.** To test whether WAI moderates the relationship between engagement and outcomes, an interaction term (WAI × log[session count]) was added to the regression models. The interaction was non-significant for PHQ-9 ($\beta = -0.01$, $p = 0.901$) and GAD-7 ($\beta = -0.02$, $p = 0.752$). However, the interaction was significant for MSPSS ($\beta = 0.24$, $p = 0.001$), indicating that the positive effect of session engagement on social support was stronger among users with higher therapeutic alliance (Figure 6).

# Discussion

## Principal Findings

This study offers three main contributions to the emerging literature on AI-based mental health tools. First, engagement with conversational AI is not monolithic; users organize into qualitatively distinct phenotypes, including a novel Concentrated User pattern, suggesting that reliance on simple session counts may obscure meaningful behavioral heterogeneity. Second, clinical benefit appears to be dimension-specific: what predicts improvement depends on the outcome in question, with within-session depth tracking depression gains, distributed consistency tracking social support, and no engagement feature capturing anxiety change over the 3-week study period. This challenges the implicit assumption in DMHI evaluation that *more engagement equals better outcomes* and argues for multidimensional assessment frameworks. Third, the working alliance findings suggest that the relational mechanisms long understood to drive change in human psychotherapy may extend to AI interactions, but that they are not a prerequisite for benefit.

**Figure 6.** WAI x engagement interaction effect on social support

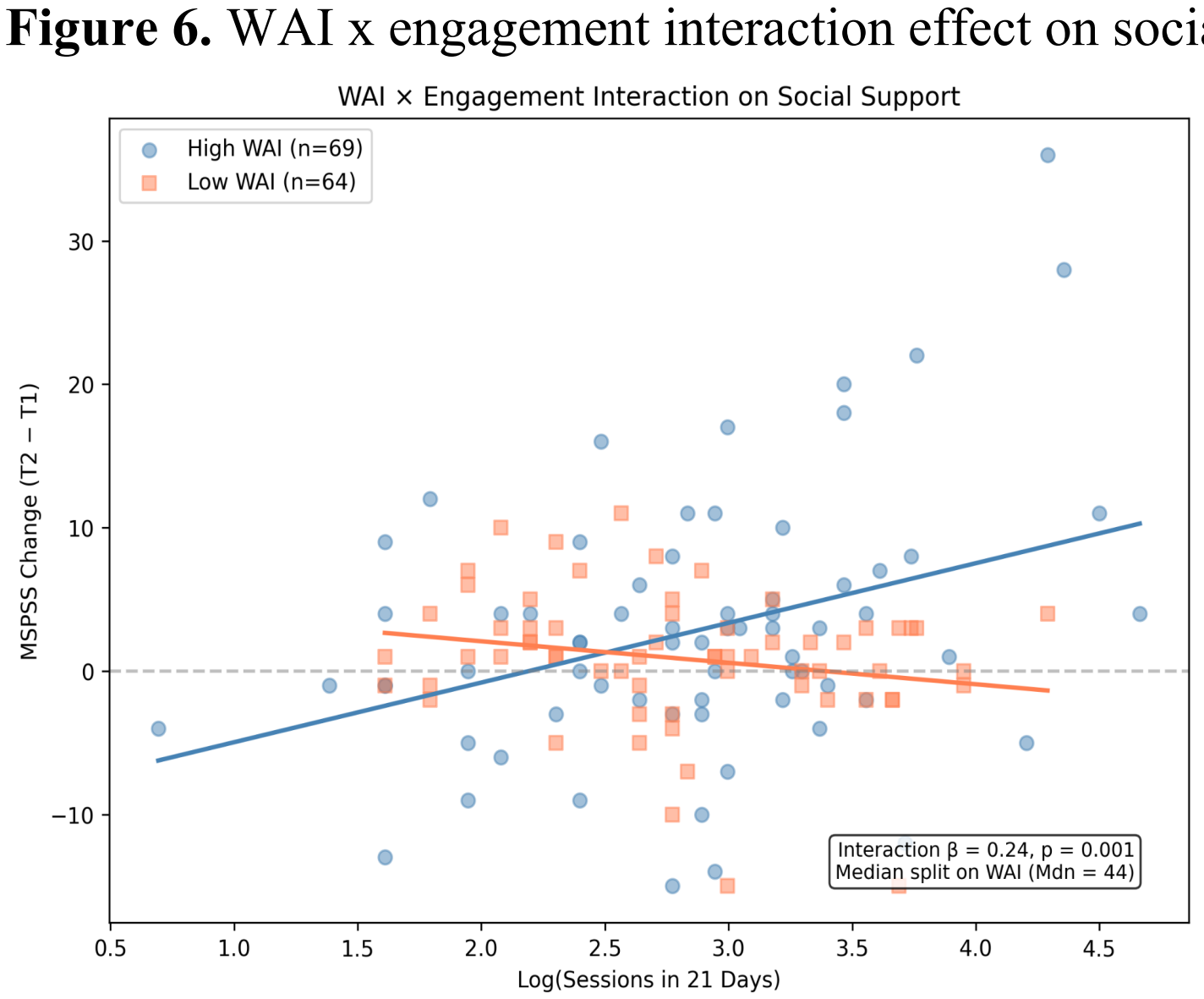

Pre-post analysis of clinical subsamples revealed significant improvements in depression, anxiety, and social support from T1 to T2. Notably, the PHQ-9 and GAD-7 differences approached but did not reach MCID thresholds in the 3-week time period. Within the PHQ-9 subsample, all four engaged clusters showed improvements in depressive symptoms, with power users showing the largest gains; though no significant self-report differences emerged between clusters, cluster-wise differences were significant with estimated PHQ-9 scores in the larger, more powered sample ($n$ = 23,813). Within the GAD-7 subsample, improvements in anxiety were significant for power users, weekly users, and concentrated users, but not for intensive users, with no significant differences between the former three clusters. No single cluster was associated with increased social support in the MSPSS subsample, but active span (e.g. the number of days spanned between a user's first and last session) significantly predicted improvement in social support from T1 to T2.

WAI scores differed significantly across engagement clusters. Power Users (C1) reported the highest alliance scores ($M = 45.1$), followed by Intensive Users (C2; $M = 44.4$) and Weekly Users (C3; $M = 44.3$). Early Dropouts (C0; $M = 42.2$) and Concentrated Users (C4; $M = 42.4$) reported the lowest WAI scores. Correlational analyses in the full WAI sample confirmed small but significant positive associations between WAI and engagement quantity (based on session count, active span, and session duration). Finally, in the clinical subsample with WAI data (n = 209), higher baseline WAI significantly predicted greater PHQ-9 improvement, but not greater GAD-7 improvement.

**Defining Engagement Clusters**

The Early Dropout cluster (C0; 52.2% of users) represents the modal use pattern for consumer-facing AI mental health tools [14,46], a rate generally lower than attrition observed in other real-world DMHI trials.[18] While duration of use can be associated with improved outcomes, recent research challenges the assumption that disengagement uniformly reflects failure; users have demonstrated benefit from as few as 1-2 sessions with purpose-built chatbots.[17,20,22,47,48] Whether early disengagement reflects unmet need or successful brief use remains an important question for future research.

The four emergent clusters map onto previously identified DMHI typologies: Power Users correspond to the well-documented high-engagement group;[19,22–24] Intensive Users, with their sporadic but concerted use within a discrete time period, resemble “deep” users;[22] Weekly Users resemble a group identified as “light” users.[22,23] Beyond these established categories, the present work suggests that Power Users of chatbots may also be characterized by high consistency of use (i.e., contiguous strings of active days), a dimension not typically captured in prior engagement research. The fifth and second-largest phenotype, Concentrated Users, appears novel to our findings. These users compress substantial engagement into roughly one week and then disengage, sharing more in common with Early Dropouts in total exposure than with sustained-use clusters, but differing meaningfully on outcomes.

**Associations Between Engagement Clusters and Dose Response**

User-reported reductions in depression and anxiety symptoms and improvements in social support represent an early, encouraging sign of clinical impact in a real world, unselected, uncompensated, free-use sample. The directional gradient in self-reported PHQ-9 improvement, with Power Users showing the largest effect sizes, suggests a potential dose-response relationship that the small self-report subsample may have been underpowered to detect. In the larger predicted-PHQ sample (n = 23,813), a similar gradient emerged: the high-engagement phenotypes improved most, while episodic users improved roughly half as much, with between-cluster differences reaching significance. That distinct engagement styles appeared to reach the same benefit, rather than benefit scaling linearly with session frequency, suggests that multiple engagement styles may be associated with positive outcomes.

The continuous predictor analyses add color to the cluster-level associations between engagement and outcomes. Message volume predicted PHQ-9 improvement while session count did not, suggesting that within-session depth may matter more than how often a user returns. This distinction holds implications for how DMHIs, especially LLM-based chatbots, are evaluated. For social support, consistency rather than depth was associated with improvement: active span predicted MSPSS gains, aligning with Power Users being the only cluster with significant within-cluster MSPSS improvement. No engagement metric or alliance measure predicted GAD-7 change, suggesting anxiety reduction may depend on factors not captured in this study. It is also worth acknowledging that, within this sample, there was no evidence associating very high levels of use (measured across any of the eight-engagement metrics) with diminishing returns or a negative impact on anxiety, depression, or social support. While a 21-day window does not account for long term, sustained use (e.g. over multiple months or years) this provides preliminary evidence that purpose-built chatbots may buffer against negative mental health effects associated with reported over-reliance among some users of general-purpose agents.[49–51]

Notably, one behavioral pattern that remained consistent across engagement clusters was time of use. Night-time use (9 PM-5 AM) was substantial across all clusters (66.9% of users had at least one overnight session; mean 38.6% of sessions per user). This pattern underscores the role of conversational AI as a resource during hours when traditional mental health services are unavailable regardless of other engagement characteristics. While more research is needed to understand the relative benefits and drawbacks of after-hours usage among diverse clinical populations, this finding is consistent with reported structural barriers to in-person therapy among chatbot users, many of whom disengage from or do not access therapy due to challenges with access and timing[52].

**Working Alliance and Clinical Benefit**

That users formed a measurable working alliance with Ash at all, let alone at levels comparable to those reported in face-to-face therapy [53], challenges the assumption that human-AI mental health interactions are inherently transactional. Further, echoing one of the most robust findings in the psychotherapy literature [53], alliance with the AI predicted depression improvement above and beyond baseline severity, replicating previous findings[15]. This suggests that the relationship between working alliance and clinical change may not be unique to human-human services. The divergence for GAD-7 hints that this relational pathway was not captured in this study.

Alliance moderated the session-outcome relationship for MSPSS but not for PHQ-9 or GAD-7: more frequent engagement translated into larger social support gains among users with stronger alliance. The null moderation for symptom outcomes suggests the clinical benefit of Ash is robust to variation in alliance: both high- and low-alliance users showed comparable PHQ-9 and GAD-7 reductions, indicating that meaningful improvement does not *require* users to form a strong relational bond with the AI.

**Strengths, Limitations, and Future Directions**

*Strengths and limitations.* Strengths of the study include a large, real-world sample diverse with respect to age and racial identity, whose participation was entirely self-guided and uncompensated, and a multidimensional characterization of engagement beyond session counts. Several limitations temper interpretation. First, clinical subsamples were modest ($n = 194$-$298$) and unevenly distributed across clusters. Adequate power ($n \geq 30$) was available for Power Users (C1), Weekly Users (C3), and Concentrated Users (C4); Early Dropouts (C0) and Intensive Users (C2) were substantially underpowered, such that cluster-level results should be interpreted cautiously. Second, engagement features capture how much and how often users engaged but not what they discussed or how meaningfully they interacted with Ash's responses, suggesting value in future work on content-level analyses to differentiate effective from ineffective engagement.

*Future directions.* Certain behavioral characteristics (e.g. night time/after hours user) were highly consistent across clusters, suggesting a need to understand how chatbots may fill gaps in available social connection and/or professional services. The small but significant improvements in social support reported by users in the clinical subsample provide evidence that purpose-built design may mitigate potentially isolating effects; more work is needed directly comparing outcomes with purpose-built versus general AI models for emotional support. Further, future research could explore the observed association between greater active span and increased social support from T1 to T2, which—in line with prior research on purpose-built models[15]—suggests that some element of consistency may be key to generalizing relational benefit to real-world settings. Finally, more work is needed to explicate the relationship between engagement, working alliance, and outcomes, exploring how and for whom working alliance with AI plays a role in clinically significant change.

**Conclusions**

This study provides one of the largest naturalistic characterizations to date of real-world engagement with a purpose-built conversational AI for mental health. Among 102,684 users, we identified five engagement phenotypes, including a novel Concentrated User pattern, and found that meaningful clinical improvement occurred across a heterogeneous, unselected user base. Different outcomes were associated with different dimensions of engagement, and working alliance contributed to depression improvement above and beyond engagement quantity without being a prerequisite for benefit. Together, these findings offer encouraging real-world evidence for the clinical value of purpose-built conversational AI as a scalable mental health resource, while highlighting the need for content-level, longer-trajectory research to clarify how engagement quality and individual differences shape who benefits most.

# Supplementary Materials

## 1. Supplementary Methods

### 1.1. Engagement feature data cleaning

Session duration was capped at a maximum of 120 minutes per session to reduce the influence of sessions where the app was left open without active use. Session count and related features showed right-skewed distributions (skewness = 2.92 for raw session count); log-transformations were applied to session count in the dose-response regression models. Active span days showed near-normal distribution (skewness = 0.19) and was not transformed.

### 1.2. Clinical Outcome Data

Clinical outcome data were distributed unevenly across clusters, reflecting differential rates of in-app survey completion (Table 3; C3: Weekly Users contributed 62.1% of the clinical subsample, while C0: Early Dropouts contributed only 4.0%). Adequate power ($n \geq 30$) was available for Power Users (C1), Weekly Users (C3), and Concentrated Users (C4); Early Dropouts (C0) and Intensive Users (C2) were substantially underpowered and results should be interpreted cautiously.

### 1.3. Analytic methods: PHQ-9 prediction

To supplement self-reported PHQ-9 assessments, we used an ensemble of five fine-tuned large language models to predict PHQ-9 total scores directly from conversation transcripts (full methodological details are reported in Tieleman et al. [2026]). Briefly, each ensemble member was built on a continued-pretraining variant of Qwen3.5-27B adapted to Ash conversations, fine-tuned with LoRA (low-rank adaptation; Hu et al., 2021) and a mixed head predicting the PHQ-9 total and its nine items; the five members differed in training-data composition and hyperparameters, and per-window predictions were averaged across them. Members were trained end-to-end (MSE loss on the regression target) on three sources: conversations with ground-truth PHQ-9 assessments collected at intake; harvested high-confidence pseudo-labels (added to offset the skew toward higher PHQ-9 scores in the ground-truth data); and a real-world prospective cohort contributing administered PHQ-9 and GAD-7 scores, which improved calibration in the minimal-to-mild range.

On a held-out validation set combining 743 original ground-truth windows with 541 windows from a real-world prospective cohort, the five-model ensemble achieved MAE = 4.0 on the ground-truth windows (4.2 across all windows) and AUC = 0.84 for the clinical case threshold (PHQ-9 ≥ 10); predicted scores correlated with administered PHQ-9 at Pearson $r$ = 0.64 ($R^2$ = 0.41). Accuracy was strongest in the moderate range (per-band MAE 3.6-3.8) and weaker at the floor and ceiling. Restricting to the most confident 50% of windows improved accuracy markedly (MAE = 3.71, AUC ≥ 10 = 0.91, AUC ≥ 20 = 0.85), motivating the high-confidence sensitivity analyses reported in the manuscript.

**1.4. Analytic methods: Individual engagement features and clinical outcomes**

As a complement to analyses examining clinical outcome change across engagement clusters, we examined relationships between raw engagement features and outcomes. To test the prospective relationship between early engagement and clinical outcomes, all engagement metrics for the dose-response analysis were restricted to the first 21 days of app use, computed using exact session and message timestamps from the production database. This 21-day window was selected to align with the three-week interval between clinical assessment timepoints, ensuring that the engagement predictor temporally preceded the outcome measurement. Ordinary least squares (OLS) regression models were estimated predicting time 3 outcome score, controlling for baseline score, to examine unique associations beyond regression to the mean.

## 2. Supplementary Results

### 2.1. Sensitivity Analysis: 21-Day Engagement Features as Predictors of Clinical Change

After controlling for baseline PHQ-9 severity, three 21-day engagement features significantly predicted week-3 depression scores. Total message volume showed the strongest effect ($\beta$ = -0.14, *SE* = 0.05, $sr^2$ = 0.031, *p* = 0.009), followed by average messages per session ($\beta$ = -0.12, *SE* = 0.05, $sr^2$ = 0.023, *p* = 0.022) and total active messaging time ($\beta$ = -0.12, *SE* = 0.05, $sr^2$ = 0.021, *p* = 0.029). Session count showed a trending effect ($\beta$ = -0.09, *SE* = 0.05, *p* = 0.092). Average active messaging time, sessions per week, and active span were not significant ($p > .24$; Supp. Table 3). No 21-day engagement features significantly predicted week 3 GAD-7 scores after controlling for baseline (all $p > .28$; Supp. Table 3). Two engagement features significantly predicted improved week-3 social support, controlling for baseline: Total active messaging time ($\beta$ = 0.17, SE = 0.06, $sr^2$ = 0.041, *p* = 0.005) and session count ($\beta$ = 0.12, *SE* = 0.06, $sr^2$ = 0.021, *p* = 0.045). Total messages ($\beta$ = 0.11, *p* = 0.061) and active span days ($\beta$ = 0.10, *p* = 0.082) showed trending effects. Average messages per session, average active messaging time, and sessions per week were not significant ($p > .22$; Supp. Table 3).

## 3. Supplementary Tables

### 3.1. Supp. Table 1. *Baseline-to-Week 3 Change on PHQ-9, GAD-7, and MSPSS*

| Outcome | *n* | Baseline *M* (*SD*) | Week 3 *M* (*SD*) | Change *M* (*SD*) | *t* | Cohen's *d* | *p* |
|---|---|---|---|---|---|---|---|
| PHQ-9 | 298 | 16.31 (6.12) | 13.22 (6.94) | -3.08 (6.04) | -8.81 | -0.55 | < .001 |
| GAD-7 | 298 | 11.90 (5.08) | 9.12 (5.18) | -2.79 (4.91) | -9.80 | -0.59 | < .001 |
| MSPSS | 194 | 24.96 (7.57) | 26.54 (8.18) | +1.57 (7.18) | +3.05 | +0.21 | .002 |

**Note.** Each outcome was measured in an independent, non-overlapping subsample (zero users shared across groups). PHQ-9: *n* = 298. GAD-7: *n* = 298. MSPSS: *n* = 194. *t* = paired-samples t-test statistic; df = *n* - 1. Cohen's d computed using pre-post SD.

**3.2. Supp. Table 2.** Baseline-to-Follow-Up Outcomes by Engagement Cluster (Predicted PHQ)

| Cluster | *n* | Baseline *M* (*SD*) | Follow-up *M* (*SD*) | Change *M* (*SD*) | Cohen's *d* | *p* |
|---|---|---|---|---|---|---|
| C0: Early dropouts | 989 | 12.98 (5.43) | 12.43 (5.49) | -0.55 (4.99) | -0.11 | < .001 |
| C1: Power users | 1,611 | 13.35 (5.65) | 11.29 (5.67) | -2.06 (5.13) | -0.40 | < .001 |
| C2: Intensive users | 1,703 | 14.24 (5.59) | 12.15 (5.59) | -2.08 (4.84) | -0.43 | < .001 |
| C3: Weekly users | 17,048 | 13.02 (5.42) | 11.85 (5.33) | -1.16 (5.06) | -0.23 | < .001 |
| C4: Concentrated users | 2,462 | 13.39 (5.57) | 11.70 (5.37) | -1.69 (4.99) | -0.34 | < .001 |
| *Overall* | 23,813 | 13.16 (5.48) | 11.84 (5.39) | -1.32 (5.05) | -0.26 | < .001 |

**Note.** One-way ANOVA across clusters, $F(4, 23{,}808) = 31.49$, $p < .001$. $M$ = mean; $SD$ = standard deviation. Cohen's $d$ = change $M$ / change $SD$ (paired); negative values indicate symptom reduction. Cluster-level $p$ values are from paired-samples $t$ tests of baseline-to-follow-up change.

**3.3. Supp. Table 3. Predicted PHQ-9 change by cluster (Cohen's d): robustness across time horizons and confidence subsets**

| **Cluster** | **First-to-Last, Full (n=23,813)** | **First-to-Last, High-conf. (n=12,122)** | **4-Week, Full (n=9,282)** | **4-Week, High-conf. (n=4,947)** |
|---|---|---|---|---|
| C0: Early Dropouts | -0.11 | -0.17 | -0.15 | -0.15 |
| C1: Power Users | -0.40 | -0.41 | -0.42 | -0.42 |
| C2: Intensive Users | -0.43 | -0.41 | -0.40 | -0.43 |
| C3: Weekly Users | -0.23 | -0.25 | -0.28 | -0.29 |
| C4: Concentrated Users | -0.34 | -0.36 | -0.34 | -0.39 |
| One-way ANOVA F | F(4,23808)=31.49 | F(4,12117)=12.84 | F(4,9277)=7.05 | F(4,4942)=4.53 |

**Note.** Updated five-model ensemble. Negative d indicates symptom improvement; all within-cluster changes p < .001 (p = .0012 for the 4-week high-confidence ANOVA). High-engagement clusters (Power, Intensive, Concentrated) showed the largest improvements across all four analyses.

**3.4. Supp. Table 4.** *Regression Models: 21-Day Engagement Predicting Clinical Change*

| Engagement feature | PHQ-9 (*n* = 224) | GAD-7 (*n* = 224) | MSPSS (*n* = 194) |
|---|---|---|---|
| | β (*SE*) | β (*SE*) | β (*SE*) |
| log(n_sessions) | -0.09 (0.05) † | -0.02 (0.06) | 0.12 (0.06) * |
| log(total_messages) | -0.14 (0.05) ** | 0.01 (0.06) | 0.11 (0.06) † |
| log(avg_messages_per_session) | -0.12 (0.05) * | 0.05 (0.06) | 0.06 (0.06) |
| total_active_messaging_time | -0.12 (0.05) * | 0.06 (0.06) | 0.17 (0.06) ** |
| avg_active_messaging_time | -0.03 (0.05) | 0.06 (0.06) | 0.04 (0.06) |
| log(sessions_per_week) | -0.06 (0.05) | 0.02 (0.06) | 0.07 (0.06) |
| active_span_days | -0.06 (0.06) | -0.06 (0.06) | 0.10 (0.06) † |

**Note.** Each row shows the standardized regression coefficient (β) and standard error from a separate model predicting Week 3 scores from the engagement feature, controlling for baseline. A negative β for PHQ-9/GAD-7 indicates that more engagement predicts lower (better) symptoms; a positive β for MSPSS indicates that more engagement predicts higher (better) perceived social support. All engagement variables are restricted to the first 21 days of use.

† $p < .10$. * $p < .05$. ** $p < .01$. *** $p < .001$.

**3.5. Supp. Table 5.** WAI Predicting Week-3 Outcomes Controlling for Baseline

| Outcome | *n* | β | *SE* | *t* | $sr^2$ | *p* | 95% CI |
|---|---|---|---|---|---|---|---|
| PHQ-9 | 209 | -0.13 | 0.06 | -2.39 | .027 | .018 * | [-0.24, -0.02] |
| GAD-7 | 207 | -0.09 | 0.06 | -1.55 | .012 | .122 | [-0.21, 0.02] |
| MSPSS | 133 | 0.18 | 0.07 | 2.49 | .045 | .014 * | [0.04, 0.32] |

**Note.** Each row reports a separate hierarchical regression in which the corresponding baseline score was entered first and the Working Alliance Inventory (WAI) total score was entered second. β is the standardized regression coefficient for WAI, *SE* its standard error, and $sr^2$ the squared semipartial correlation. PHQ-9 = Patient Health Questionnaire-9; GAD-7 = Generalized Anxiety Disorder-7; MSPSS = Multidimensional Scale of Perceived Social Support; CI = confidence interval. * $p < .05$.

## 4. Supplementary Figures

### 4.1. Supp. Figure 1. Engagement cluster profiles across five clusters.

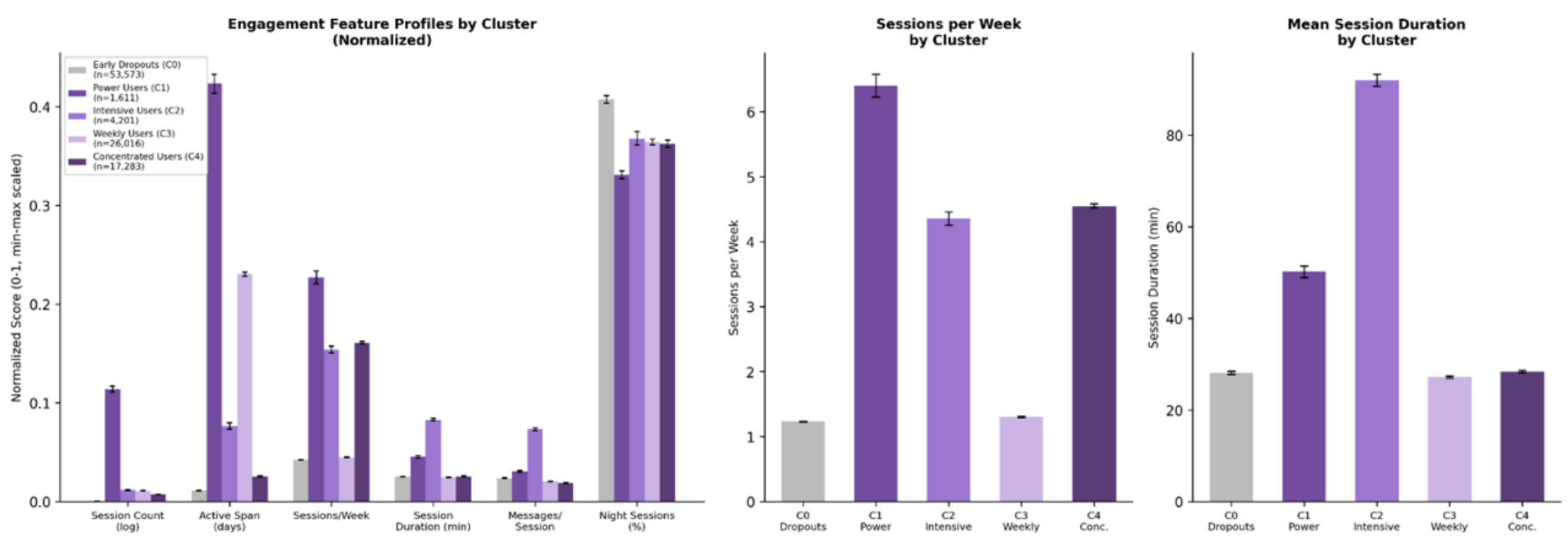

**Supp. Figure 1 Caption:** Left: Grouped bar chart showing normalized engagement features (min-max scaled to sample range) per cluster. Middle: Sessions per week by cluster. Right: Average session duration by cluster.

### 4.2. Supp. Figure 2: Relationship between log-transformed message volume and PHQ-9 change (a) and between log-transformed session volume and MSPSS change (b).

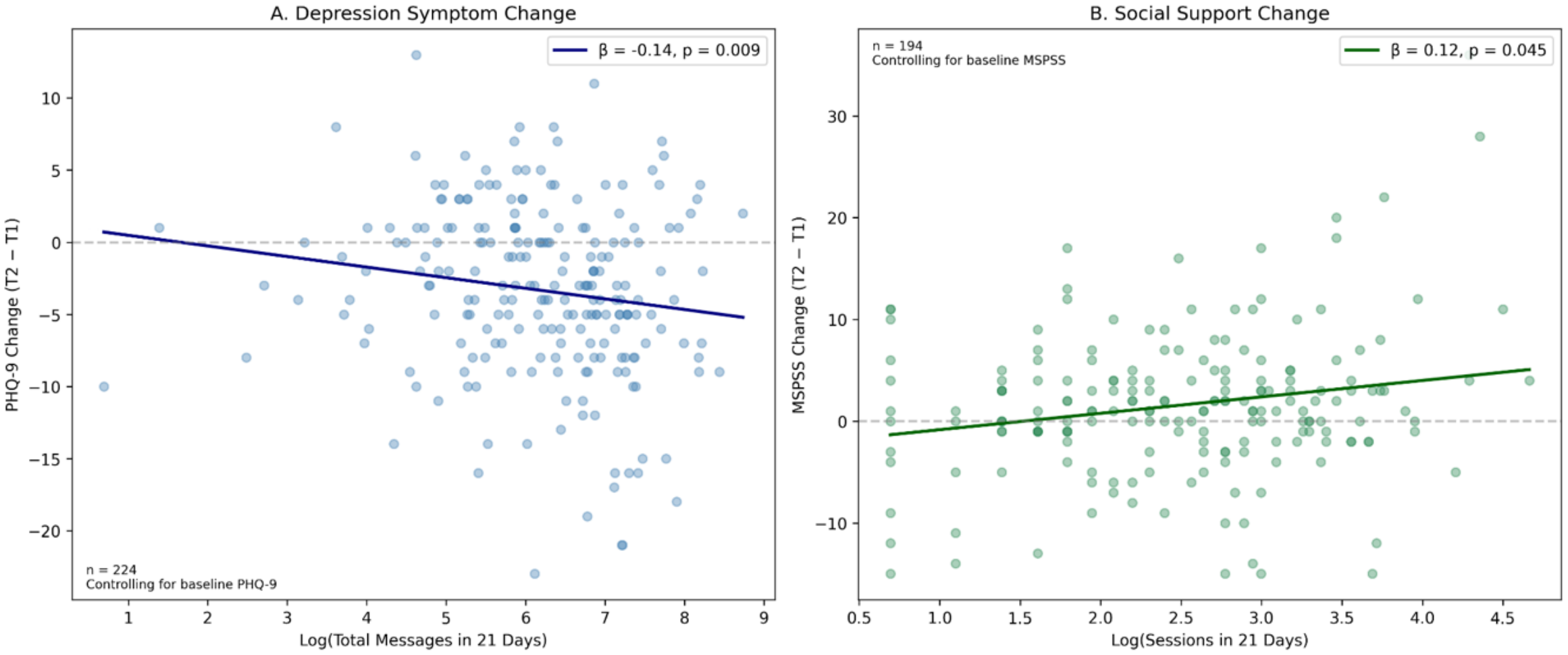